\documentstyle[12pt]{article}
\begin{document}
{\pagestyle{empty}
\rightline{TU-08/96}
\rightline{Aug. 1996}
\rightline{~~~~~~~~~}
\vskip 1cm
\centerline{\Large \bf Brans-Dicke Theory on $M_4\times Z_2$ Geometry}
\vskip 1cm

\centerline{
  {Akira Kokado\footnote{E-mail address: kokado@kgupyr.kwansei.ac.jp}},
  {Gaku Konisi\footnote{E-mail address: konisi@kgupyr.kwansei.ac.jp}},
  {Takesi Saito\footnote{E-mail address: tsaito@jpnyitp.yukawa.kyoto-u.ac.jp}}
  and 
  {Kunihiko Uehara\footnote{E-mail address: uehara@tezukayama-u.ac.jp}}}
\vskip 1cm
\centerline{\it Kobe International University, 
  Kobe 655, Japan${}^1$}
\centerline{\it Department of Physics, Kwansei Gakuin University, 
  Nishinomiya 662, Japan${}^{2,3}$}
\centerline{\it Department of Physics, Tezukayama University, 
  Nara 631, Japan${}^4$}

\vskip 2cm

\centerline{\bf Abstract}
\vskip 0.2in

  The gauge theory on $M_4\times Z_2$ geometry is applied to the 
Brans-Dicke(BD) theory, where $M_4$ is the four dimensional space-time 
and $Z_2$ is a discrete space with two points.  
This approach had been previously proposed by Konisi and Saito 
without recourse to noncommutative geometry(NCG).  
Since our approach is geometrically simpler and clearer than NCG, 
one can see more directly the effect of the $Z_2$ space in 
obtaining the BD theory.

\vskip 0.4cm\noindent
PACS number(s): 
\hfil
\vfill
\newpage}
%%%%%%%%%%%%%%%%%%%%%%%%%%%%%% Section 1 %%%%%%%%%%%%%%%%%%%%%%%%%%%%%%%
\newcommand{\iDelta}{{\mit\Delta}}
\newcommand{\iGamma}{{\mit\Gamma}}
\renewcommand{\theequation}{\thesection.\arabic{equation}}
\setcounter{equation}{0}
\addtocounter{section}{1}
\section*{\S\thesection.\ \ Introduction}

\indent

  The noncommutative geometry(NCG) of Connes\cite{Connes1,Connes2} has been 
successful in giving a geometrical interpretation of the standard model 
as well as some grand unification models.  
In this interpretation the Higgs fields are regarded as gauge fields 
along directions in the discrete space.  
The bosonic parts of actions are just the pure Yang-Mills actions 
containing gauge fields on both continuous and discrete spaces.  
They contain Higgs potentials which automatically generate the spontaneous 
symmetry breaking.  The Yukawa coupling is regarded as a kind of gauge 
interactions of fermions.

  There are now various alternative versions of NCG\cite{NCG}.  
Any NCG, however, has so far been algebraic rather than geometric.  
Nobody has considered the original geometric meaning such as covariant 
differences, parallel transportations of vector fields, curvatures and 
so on in the discrete space.  
In a previous paper\cite{Konisi} two of the authors(G.K. and T.S.) 
have considered such a geometric meaning of NCG and proposed a gauge 
theory on $M_4\times Z_N$ geometry without recourse to any knowledge of 
such NCG.  
This approach appears to be geometrically much simpler and clearer than NCG.  
Here the Higgs fields are introduced as mapping functions between 
any pair of vector fields belonging independently to the $N$-sheeted 
space-time, just as the Yang-Mills fields are so between both vectors on 
$x$ and $x+\iDelta x$.  
This approach has been applied to the Weinberg-Salam model for 
electroweak interactions\cite{Konisi} and to the $N=2$ and $4$ super 
Yang-Mills theories\cite{Konisi}.

  Besides model buildings in particle physics there are some works on gravity 
in terms of NCG on $M_4\times Z_2$\cite{Chamseddine}--\cite{Konisi2}, 
where a scalar field is coupled to gravity.  
However, the scalar fields in those papers are different from each other.
The reason of this owes partly to different torsion-free conditions used 
in their papers.  
The so-called ``partial'' torsion-free condition used in 
Refs.\cite{Chen,Konisi2} exactly leads to the Brans-Dicke(BD) 
theory\cite{Brans}.
However, the use of ``total'' torsion-free condition makes NCG into the usual 
commutative geometry and leads to the other kinds of scalar-tensor 
theory (including electromagnetism of the Kaluza-Klein type).

  Another reason owes mostly to the lack of a proper formalism of gravity 
in terms of NCG.  
In this paper we consider the derivation of the BD theory from the gauge  
theory on $M_4\times Z_2$ without recourse to NCG.  
For this purpose we need the equivalence assumption proposed in 
the previous work\cite{Konisi2}.  
This is stated as follows:  The manifold $M_4\times Z_2$ may be regarded as 
a pair of the $M_4$'s, each at the point $p=+$ or $p=-$ on $Z_2$.
The physics requires that these two $M_4$-pieces should be equivalent to 
each other.  
We assume that the equivalence is attained by a limiting process with some 
parameter $\varepsilon$ which tends to zero.  
As a technical tool to express the limiting process we introduce a coordinate 
$z(p)$ on $Z_2(p=\pm)$ such that a difference
\begin{eqnarray}
\iDelta z(p)=z(p)-z(p')\sim\varepsilon\ \ \ (p\ne p')
\label{e101}
\end{eqnarray}

\noindent
is proportional to the limiting process parameter $\varepsilon$. 
Let $\psi(z(p))$ be any function on $Z_2$.  
Since $\psi(z(p))$ is a linear function of $z(p)$, one may put 
\begin{eqnarray}
\psi(z(p))=A+Bz(p),
\label{e102}
\end{eqnarray}

\noindent
where $A$ and $B$ are some constants.  
From this we find that its Taylor expansion is cut off only up to the 
first order $\iDelta z(p)$ 
\begin{eqnarray}
\psi(z(p))=\psi(z(p'))+B\iDelta z(p).
\label{e103}
\end{eqnarray}

\noindent
Now, by the equivalence we mean that 
\begin{eqnarray}
\psi(x,z(p))\rightarrow \psi(x,z(p'))\ \ {\rm as}\ \ \varepsilon\rightarrow 0,
\label{e104}
\end{eqnarray}

\noindent
where the coordinate $x$ on $M_4$ is inserted.

  In the next section, by using this equivalence assumption we derive 
the BD theory.  
Since our approach is geometrically simple, one can see more directly 
the effect of the $Z_2$ space.  
The final section is devoted to concluding remarks.

%%%%%%%%%%%%%%%%%%%%%%%%%%%%%% Section 2 %%%%%%%%%%%%%%%%%%%%%%%%%%%%%%%
\setcounter{equation}{0}
\addtocounter{section}{1}
\section*{\S\thesection.\ \ Brans-Dicke Theory on $M_4 \times Z_2$}

\indent

  We regard the manifold $M_4\times Z_2$ to be the Kaluza-Klein like space 
where the fifth continuous dimension is replaced by two discrete points 
$z(+)$ and $z(-)$.  
The line element $\iDelta s$ of this space is assumed to be 
\begin{eqnarray}
\iDelta s^2&=&g_{\mu\nu}(x)\iDelta x^\mu\iDelta x^\nu+\lambda^2(x)\iDelta z^2
\nonumber\\
          &=&G_{MN}(x)\iDelta x^M\iDelta x^N,
\label{e201}
\end{eqnarray}

\noindent
where 
\begin{eqnarray}
\iDelta x^N=(\iDelta x^\mu,\iDelta x^5\equiv\iDelta z),\ \ \mu=0,1,2,3
\label{e202}
\end{eqnarray}

\noindent
and $\iDelta z=z(+)-z(-)$.  
Here we have considered a simple case that the four-dimensional 
metric $g_{\mu\nu}(x)$ and the scalar field $\lambda(x)$ are 
independent of $z(\pm)$ on $Z_2$ and are functions only of $x\in M_4$.  
The $G_{MN}(x)$ is regarded as the five-dimensional metric of 
$M_4\times Z_2$.  

  Let $\psi^M(x,p)$ be a vector field on $M_4\times Z_2$ with $p=(+,-)$.  
We first consider the parallel transportation of $\psi^M(x,p)$ from $(x,p)$ 
to $(x+\iDelta x,p)$.  
As usual this is achieved by the mapping function 
$H^M_{\,N}(x+\iDelta x,x,p)$ 
\begin{eqnarray}
\psi^M_\parallel(x+\iDelta x,p)=H^M_{\,N}(x+\iDelta x,x,p)\psi^N(x,p).
\label{e203}
\end{eqnarray}

\noindent
In the familiar notation with the affine connection 
$\iGamma^M_{\,\,N\mu}(x)$ we set 
\begin{eqnarray}
H^M_{\,N}(x+\iDelta x,x,p)=\delta^M_{\,N}
              -\iGamma^M_{\,\,N\mu}(x)\iDelta x^\mu
              +\frac{1}{2}C^M_{\,\,N\mu\nu}(x)\iDelta x^\mu\iDelta x^\nu
              +\cdots,
\label{e204}
\end{eqnarray}

\newlength{\minitwocolumn}
\setlength{\minitwocolumn}{.5\textwidth}
\addtolength{\minitwocolumn}{-.5\columnsep}
\ \\
\begin{minipage}[t]{\minitwocolumn}
where $\iGamma^M_{\,\,N\mu}(x)$ and $C^M_{\,\,N\mu\nu}(x)$ are assumed to be 
independent of $p=\pm\in Z_2$.  
We then consider the Riemann curvature of the usual type corresponding 
to Fig.1.  
The parallel transportations of $\psi^M(x,p)$ from $x$ to 
$x+\iDelta_1 x+\iDelta_2 x$ through $x+\iDelta_1 x$ is given by 
\end{minipage}
\hspace{\columnsep}
\begin{minipage}[t]{\minitwocolumn}
\begin{center}
\ \vspace{0ex}\\
\setlength{\unitlength}{1mm}
\begin{picture}(80,30)(-6,-5)
 \put(30,5){\vector(-1,1){6}}\put(24,11){\line(-1,1){4}}
 \put(20,15){\vector(1,1){6}}\put(26,21){\line(1,1){4}}
 \put(30,5){\vector(1,1){6}}\put(36,11){\line(1,1){4}}
 \put(40,15){\vector(-1,1){6}}\put(34,21){\line(-1,1){4}}
 \put(30,5){\circle*{1}}
 \put(20,15){\circle*{1}}
 \put(40,15){\circle*{1}}
 \put(30,25){\circle*{1}}
 \put(19,27){$x+\iDelta_1 x+\iDelta_2 x$}
 \put(4,14){$x+\iDelta_1 x$}
 \put(42,14){$x+\iDelta_2 x$}
 \put(29,1){$x$}
 \put(19,20){$C_1$}
 \put(36,20){$C_2$}
 \put(5,1){{\bf Fig.1.}}
\end{picture}
\vspace{0ex}\\
\end{center}
\end{minipage}
\begin{eqnarray}
C_1=H^M_{\,L}(x+\iDelta_1 x+\iDelta_2 x,x+\iDelta_1 x)
    H^L_{\,N}(x+\iDelta_1 x,x)\psi^N(x,p).
\label{e205}
\end{eqnarray}

\noindent
In the same way the other parallel transportations along the path $C_2$ is 
\begin{eqnarray}
C_2=H^M_{\,L}(x+\iDelta_1 x+\iDelta_2 x,x+\iDelta_2 x)
    H^L_{\,N}(x+\iDelta_2 x,x)\psi^N(x,p).
\label{e206}
\end{eqnarray}

\noindent
The difference between two parallel transportations $C_1$ and $C_2$ 
gives the Riemann curvature $R^M_{\,\,N\mu\nu}$.  
Substituting (\ref{e204}) into $C_1$ and $C_2$, we have 
\begin{eqnarray}
C_1-C_2=-R^M_{\,\,N\mu\nu}(x)\psi^N(x,p)\iDelta_1 x^\mu\iDelta_2 x^\nu,
\label{e207}
\end{eqnarray}

\noindent
where 
\begin{eqnarray}
R^M_{\,\,N\mu\nu}(x)=
       \partial_\mu\iGamma^M_{\,\,N\nu}-\partial_\nu\iGamma^M_{\,\,N\mu}
       +\iGamma^M_{\,\,L\mu}\iGamma^L_{\,\,N\nu}
       -\iGamma^M_{\,\,L\nu}\iGamma^L_{\,\,N\mu}.
\label{e208}
\end{eqnarray}

\ \\
\begin{minipage}[t]{\minitwocolumn}
\hspace{5mm}
  The Riemann curvature of the second type corresponds to Fig.2.  
The parallel transportations of $\psi^M(x,+)$ from $(x,+)$ to 
$(x+\iDelta x,-)$ through $(x,-)$ and $(x+\iDelta x,+)$ are given by 
\end{minipage}
\hspace{\columnsep}
\begin{minipage}[t]{\minitwocolumn}
\begin{center}
\ \vspace{-1ex}\\
\setlength{\unitlength}{1mm}
\begin{picture}(80,30)(-10,-7)
 \put(10,5){\line(1,0){10}}
 \put(10,20){\line(1,0){10}}
 \put(35,5){\line(1,0){10}}
 \put(35,20){\line(1,0){10}}
 \put(20,5){\vector(0,1){8}}\put(20,13){\line(0,1){7}}
 \put(20,20){\vector(1,0){8}}\put(28,20){\line(1,0){7}}
 \put(20,5){\vector(1,0){8}}\put(28,5){\line(1,0){7}}
 \put(35,5){\vector(0,1){8}}\put(35,13){\line(0,1){7}}
 \put(20,5){\circle*{1}}
 \put(20,20){\circle*{1}}
 \put(35,5){\circle*{1}}
 \put(35,20){\circle*{1}}
 \put(19,1){$x$}
 \put(31,1){$x+\iDelta x$}
 \put(19,22){$x$}
 \put(31,22){$x+\iDelta x$}
 \put(50,4){$p=+$}
 \put(50,19){$p=-$}
 \put(24,22){$C_3$}
 \put(37,11){$C_4$}
 \put(-2,1){{\bf Fig.2.}}
\end{picture}
\vspace{-1ex}\\
\end{center}
\end{minipage}
\begin{eqnarray}
C_3=H^M_{\,L}(x+\iDelta x,x,-)
    H^L_{\,N}(x,-,+)\psi^N(x,+)
\label{e209}
\end{eqnarray}

\noindent
and
\begin{eqnarray}
C_4=H^M_{\,L}(x+\iDelta x,-,+)
    H^L_{\,N}(x+\iDelta x,x,+)\psi^N(x,+),
\label{e210}
\end{eqnarray}

\noindent
respectively, where $H(x,-,+)$ and $H(x+\iDelta x,-,+)$ are mapping 
functions and 
\begin{eqnarray}
H^L_{\,N}(x,-,+)\psi^N(x,+)
\label{e211}
\end{eqnarray}

\noindent
shows the parallel transportation of $\psi^N(x,+)$ from $(x,+)$ to 
$(x,-)$.  
The difference between two parallel transportations $C_3$ and $C_4$ 
gives the Riemann curvature $R^M_{\,\,N\mu 5}$.  
Introducing the affine connection $\iGamma^L_{\,\,N5}(x)$, which is 
also assumed to be independent of $p$, and from the equivalence 
assumption, one may put 
\begin{eqnarray}
H^L_{\,N}(x,-,+)&=&\delta^L_{\,N}-\iGamma^L_{\,\,N5}(x)\iDelta z,
\label{e212}\\
H^L_{\,N}(x,+,-)&=&\delta^L_{\,N}+\iGamma^L_{\,\,N5}(x)\iDelta z,
\label{e213}
\end{eqnarray}

\noindent
so that
\begin{eqnarray}
C_3-C_4=R^M_{\,\,N\mu 5}(x)\psi^N(x,+)\iDelta x^\mu\iDelta z,
\label{e214}
\end{eqnarray}

\noindent
where 
\begin{eqnarray}
R^M_{\,\,N\mu 5}(x)=
       \partial_\mu\iGamma^M_{\,\,N5}
       +\iGamma^M_{\,\,L\mu}\iGamma^L_{\,\,N 5}
       -\iGamma^M_{\,\,L 5}\iGamma^L_{\,\,N\mu}.
\label{e215}
\end{eqnarray}

\noindent
Note that in Eqs.(\ref{e212}) and (\ref{e213}) there are no term of 
$O(\iDelta z^2)$ as shown in Appendix A.  
\ \\
\begin{minipage}[t]{\minitwocolumn}
\hspace{5mm}
  The Riemann curvature of the third type corresponds to Fig.3.  
Namely, $\psi^M(x,+)$ is compared with $\psi^M_\parallel(x,+)$ which is 
the parallel-transported vector of $\psi^M(x,+)$ from $(x,+)$ to $(x,-)$ 
and then returning to $(x,+)$, {\it i.e.},
\end{minipage}
\hspace{\columnsep}
\begin{minipage}[t]{\minitwocolumn}
\begin{center}
\ \vspace{-1ex}\\
\setlength{\unitlength}{1mm}
\begin{picture}(80,30)(-10,-2)
 \put(25,5){\line(1,0){10}}
 \put(25,20){\line(1,0){10}}
 \put(35,5){\line(1,0){10}}
 \put(35,20){\line(1,0){10}}
 \put(36,5){\vector(0,1){8}}\put(36,13){\line(0,1){7}}
 \put(35,20){\vector(0,-1){8}}\put(35,12){\line(0,-1){7}}
 \put(35.5,5){\circle*{1}}
 \put(35.5,20){\circle*{1}}
 \put(30,1){$(x,+)$}
 \put(30,22){$(x,-)$}
 \put(10,1){{\bf Fig.3.}}
\end{picture}
\vspace{-1ex}\\
\end{center}
\end{minipage}
\begin{eqnarray}
\psi^M(x,+)-\psi^M_\parallel(x,+)
  &=&[\,\delta^M_{\,N}-H^M_{\,L}(x,+,-)H^L_{\,N}(x,-,+)\,]\psi^N(x,+)
\nonumber\\
  &=&[\,\delta^M_{\,N}-(\delta^M_{\,L}+\iGamma^M_{\,\,L5}\iDelta z)
                   (\delta^L_{\,N}-\iGamma^L_{\,\,N5}\iDelta z)\,]\psi^N(x,+)
\nonumber\\
  &=&\iGamma^M_{\,\,L5}(x)\iGamma^L_{\,\,N5}(x)\psi^N(x,+)\iDelta z\iDelta z.
\label{e216}
\end{eqnarray}

\noindent
This gives the Riemann curvature of the third type 
\begin{eqnarray}
R^M_{\,\,N55}(x)=\iGamma^M_{\,\,L5}(x)\iGamma^L_{\,\,N5}(x).
\label{e217}
\end{eqnarray}

\noindent
On $M_4$ we know that there is no curvature of the similar type 
corresponding to two paths: 
$x\rightarrow x+\iDelta x\rightarrow x$ and $x\rightarrow x$.  
On $Z_2$, however, we find non-vanishing curvature of the third type 
(see Appendix B).

  Now we would like to express the affine connection $\iGamma^L_{\,\,MN}(x)$ 
in terms of the metric $G_{MN}(x)$.  
This is achieved in the usual way if we use the covariant-free equation 
\begin{eqnarray}
\nabla_{\!K}G_{MN}\equiv \partial_K G_{MN}
                        -\iGamma_{MKN}-\iGamma_{NKM}=0,
\label{e218}
\end{eqnarray}

\noindent
where $\iGamma_{MKN}\equiv G_{ML}\iGamma^L_{\,\,KN}$, and if 
$\iGamma_{MKN}$ is symmetric with respect to $K$ and $N$, {\it i.e.}, 
$\iGamma_{MKN}=\iGamma_{MNK}$.  
The equation (\ref{e218}) will follow from the requirement that the 
length of any vector field is invariant under the parallel transportation, 
{\it i.e.}, 
\begin{eqnarray}
G_{MN}(X+\iDelta X)\psi^M_\parallel(X+\iDelta X)\psi^N_\parallel(X+\iDelta X)
  =G_{MN}(X)\psi^M(X)\psi^N(X),
\label{e219}
\end{eqnarray}

\noindent
where $X=(x^\mu,z(p))$. 
The result is known to be
\begin{eqnarray}
\iGamma^L_{\,\,MN}=\iGamma^L_{\,\,NM}
               =\frac{1}{2}G^{LK}(\partial_M G_{KN}+\partial_N G_{KM}
                                 -\partial_K G_{MN}).
\label{e220}
\end{eqnarray}

\noindent
In the present case one may set $\partial_5 G_{MN}=0$, because $G_{MN}$ 
is independent of $z(p)$.  

  Since the metric $G_{MN}$ and its inverse $G^{MN}$ are given by 
\begin{eqnarray}
G_{MN}&=&\left(\matrix{g_{\mu\nu}(x)&0\cr
                     0&\lambda^2(x)\cr}\right),
\label{e221}\\
G^{MN}&=&\left(\matrix{g^{\mu\nu}(x)&0\cr
                     0&\lambda^{-2}(x)\cr}\right),
\label{e222}
\end{eqnarray}

\noindent
we get all components of $\iGamma^L_{\,\,MN}$
{\setcounter{enumi}{\value{equation}}
\addtocounter{enumi}{1}
\setcounter{equation}{0}
\renewcommand{\theequation}{\thesection.\theenumi\alph{equation}}
\begin{eqnarray}
\iGamma^\rho_{\,\,\mu\nu}&=&\frac{1}{2}g^{\rho\sigma}
                      (\partial_\mu g_{\sigma\nu}+\partial_\nu g_{\sigma\mu}
                      -\partial_\sigma g_{\mu\nu}),
\label{e223a}\\
\iGamma^\rho_{\,\,5\nu}&=&0,
\label{e223b}\\
\iGamma^\rho_{\,\,55}&=&-\lambda\partial^{\,\rho}\lambda,
\label{e223c}\\
\iGamma^5_{\,\,\mu\nu}&=&0,
\label{e223d}\\
\iGamma^5_{\,\,5\nu}&=&\frac{1}{\lambda}\partial_\nu\lambda,
\label{e223e}\\
\iGamma^5_{\,\,55}&=&0.
\label{e223f}
\end{eqnarray}
\setcounter{equation}{\value{enumi}}
}

\noindent
Substituting (2.23) into (\ref{e208}), (\ref{e215}) and (\ref{e217}), 
one obtains the Riemann curvature components
{\setcounter{enumi}{\value{equation}}
\addtocounter{enumi}{1}
\setcounter{equation}{0}
\renewcommand{\theequation}{\thesection.\theenumi\alph{equation}}
\begin{eqnarray}
R^\rho_{\,\,\sigma\mu\nu}&=&\partial_\mu\iGamma^\rho_{\,\,\sigma\nu}
                         -\partial_\nu\iGamma^\rho_{\,\,\sigma\mu}
          +\iGamma^\rho_{\,\,\lambda\mu}\iGamma^\lambda_{\,\,\sigma\nu}
          -\iGamma^\rho_{\,\,\lambda\nu}\iGamma^\lambda_{\,\,\sigma\mu}
          =-R^\rho_{\,\,\sigma\nu\mu},
\label{e224a}\\
R^5_{\,\,\mu 5\nu}&=&-{\mathop{\nabla}^\circ}_{\!\nu}
    (\frac{1}{\lambda}\partial_\mu\lambda)
    -\frac{1}{\lambda^2}\partial_\nu\lambda\partial_\mu\lambda
   =-\frac{1}{\lambda}{\mathop{\nabla}^\circ}_{\!\nu}\partial_\mu\lambda
   =-R^5_{\,\,\mu\nu 5},
\label{e224b}\\
R^\mu_{\,\,55\nu}&=&{\mathop{\nabla}^\circ}_{\!\nu}
    (\lambda\partial^{\,\mu}\lambda)
    -\partial^{\,\mu}\lambda\partial_\nu\lambda
   =\lambda{\mathop{\nabla}^\circ}_{\!\nu}\partial^{\,\mu}\lambda
   =-R^\mu_{\,\,5\nu 5},
\label{e224c}\\
R^\mu_{\,\,\nu 55}&=&-\partial^{\,\mu}\lambda\partial_\nu\lambda,
\label{e224d}\\
R^5_{\,\,555}&=&-\partial_\mu\lambda\partial^{\,\mu}\lambda,
\label{e224e}
\end{eqnarray}
\setcounter{equation}{\value{enumi}}
}

\noindent
and all other components are zero, where 
$\displaystyle{\mathop{\nabla}^\circ}_{\!\nu}$ 
is the covariant derivative on $M_4$, {\it i.e.},
\begin{eqnarray}
{\mathop{\nabla}^\circ}_{\!\nu}V^\mu
   &\equiv&\partial_\nu V^\mu+\iGamma^\mu_{\,\,\rho\nu}V^\rho,
\label{e225}\\
{\mathop{\nabla}^\circ}_{\!\nu}V_{\!\mu}
   &\equiv&\partial_\nu V_{\!\mu}-\iGamma^\rho_{\,\,\mu\nu}V_{\!\rho}.
\label{e226}
\end{eqnarray}

  The Ricci curvature on $M_4\times Z_2$ is, therefore, given by 
{\setcounter{enumi}{\value{equation}}
\addtocounter{enumi}{1}
\setcounter{equation}{0}
\renewcommand{\theequation}{\thesection.\theenumi\alph{equation}}
\begin{eqnarray}
R_{MN}&=&\sum_L R^L_{\,\,MLN}
\nonumber\\
      &=&R^\rho_{\,\,M\rho N}+R^5_{\,\,M5N},
\label{e227a}
\end{eqnarray}

\noindent
hence
\begin{eqnarray}
R_{\mu\nu}&=&{\mathop{R}^\circ}_{\mu\nu}+R^5_{\,\,\mu 5\nu}
\nonumber\\
          &=&{\mathop{R}^\circ}_{\mu\nu}-\frac{1}{\lambda}
             {\mathop{\nabla}^\circ}_{\!\nu}\partial_\mu\lambda,
\label{e227b}\\
R_{5\nu}&=&R_{\nu 5}=0,
\label{e227c}\\
R_{55}&=&R^\rho_{\,\,5\rho 5}+R^5_{\,\,555}
\nonumber\\
      &=&-\lambda{\mathop{\nabla}^\circ}_{\!\rho}\partial^{\,\rho}\lambda
         -\partial_\mu\lambda\partial^{\,\mu}\lambda,
\label{e227d}
\end{eqnarray}

\noindent
where
\begin{eqnarray}
{\mathop{R}^\circ}_{\mu\nu}\equiv R^\rho_{\,\,\mu\rho\nu}
\label{e227e}
\end{eqnarray}

\setcounter{equation}{\value{enumi}}
}

\noindent
is the usual four-dimensional Ricci curvature on $M_4$.  
Finally, therefore, we get the scalar curvature on $M_4\times Z_2$ 
\begin{eqnarray}
R&=&G^{MN}R_{MN}=g^{\mu\nu}R_{\mu\nu}+G^{55}R_{55}
\nonumber\\
 &=&{\mathop{R}^\circ}
  -2\frac{1}{\lambda}{\mathop{\nabla}^\circ}_{\!\mu}\partial^{\,\mu}\lambda
  -\frac{1}{\lambda^2}\partial_\mu\lambda\partial^{\,\mu}\lambda.
\label{e228}
\end{eqnarray}

\noindent
By using $\det(G_{MN})=\det(g_{\mu\nu})\lambda^2\equiv g\lambda^2$,  
we obtain the gravity action
\begin{eqnarray}
I&=&\int_{M_4}\int_{Z_2}\sqrt{-\det(G_{MN})}\,R \nonumber\\
 &=&\int_{M_4}\sqrt{-g}\lambda\,[{\mathop{R}^\circ}
  -2\frac{1}{\lambda}{\mathop{\nabla}^\circ}_{\!\mu}\partial^{\,\mu}\lambda
  -\frac{1}{\lambda^2}\partial_\mu\lambda\partial^{\,\mu}\lambda] \nonumber\\
 &=&\int_{M_4}\sqrt{-g}\,[\lambda{\mathop{R}^\circ}
  -2{\mathop{\nabla}^\circ}_{\!\mu}\partial^{\,\mu}\lambda
  -\frac{1}{\lambda}\partial_\mu\lambda\partial^{\,\mu}\lambda] \nonumber\\
 &=&\int_{M_4}\sqrt{-g}\,[\lambda{\mathop{R}^\circ}
  -\frac{1}{\lambda}\partial_\mu\lambda\partial^{\,\mu}\lambda],
\label{e229}
\end{eqnarray}

\noindent
which is just the BD theory obtained in the previous 
works\cite{Chen,Konisi2}.

%%%%%%%%%%%%%%%%%%%%%%%%%%%%%% Section 3 %%%%%%%%%%%%%%%%%%%%%%%%%%%%%%%
\setcounter{equation}{0}
\addtocounter{section}{1}
\section*{\S\thesection.\ \ Concluding remarks}

\indent

  On the basis of the equivalence assumption stated in the introduction 
we have derived the Brans-Dicke theory on the manifold $M_4\times Z_2$.  
We have used the gauge theory on $M_4\times Z_2$ geometry without recourse 
to NCG.  
The BD kinetic term comes from the Riemann curvature of the third type 
corresponding to Fig.3.  
This sharply differs from the Kaluza-Klein theory with the scalar field, 
in which there is no BD kinetic term.  

  On the continuous manifold one can see that there is no curvature 
coming from the difference between two paths: 
$x\rightarrow x+\iDelta x\rightarrow x$ and $x\rightarrow x$.  
On the discrete manifold, however, we find non-vanishing curvatures 
of the third type, which corresponds to two paths: 
$A\rightarrow B\rightarrow A$ and $A\rightarrow A$.  
These have been discussed in Appendix B.  
The result shows that the BD scalar field is related to the $Z_2$ discrete 
space and is interpreted to describe the distance between the two points 
in this space.
The affine connection $\iGamma^M_{\,\,N5}(x)$ has been introduced as a 
gauge field along the $Z_2$ direction.  
Its non-zero components are given in (\ref{e223c},e) in terms of the 
BD scalar field.

  Finally we consider the torsion of $M_4\times Z_2$.
There are three kinds of torsions: 
$\iGamma^M_{\,\,\mu\nu}-\iGamma^M_{\,\,\nu\mu}$,
$\iGamma^M_{\,\,\mu 5}-\iGamma^M_{\,\,5\mu}$ and 
$\iGamma^M_{\,\,55}$\cite{Konisi2}.  
Since the affine connection $\iGamma^M_{\,\,LN}$ is symmetric with respect 
to $L$ and $N$ , the first and second torsions vanish.  
However, the last torsion 
$\iGamma^M_{\,\,55}=-\delta^M_{\,\mu}\lambda\partial^{\,\mu}\lambda$ is not 
zero except for $\lambda={\rm constant}$.  
This is undoubtedly related to the $Z_2$ space and $\lambda$.

%%%%%%%%%%%%%%%%%%%%%%%%%%%%%% Appendix A %%%%%%%%%%%%%%%%%%%%%%%%%%%%%%%
\newpage
\renewcommand{\theequation}{\Alph{section}.\arabic{equation}}
\setcounter{equation}{0}
\setcounter{section}{1}
\section*{Appendix \Alph{section}}

\indent

  Since any function on $Z_2$ is a linear function of coordinates $z(+)=z$ 
and $z(-)=z'$, one may put
\begin{eqnarray}
H(z,z')=A+Bz+B'z'+Czz',
\label{eA01}
\end{eqnarray}

\noindent
where $A,B,B'$ and $C$ are independent of $z$ and $z'$.  
From $H(z,z)=1$ one gets
\begin{eqnarray}
A+(B+B')z+Cz^2=1.
\label{eA02}
\end{eqnarray}

\noindent
Without loss of generality we can put $z=\pm\varepsilon$, where 
$\varepsilon$ is the limiting process parameter (\ref{e101}).  
Then it follows that
\begin{eqnarray}
A=1-C\varepsilon^2,\ \ \ B'=-B,
\label{eA03}
\end{eqnarray}

\noindent
hence
\begin{eqnarray}
H(z,z')&=&1+B(z-z')+C(zz'-z^2) \nonumber\\
       &=&1+(B-Cz)\iDelta z \nonumber\\
       &=&1+\iGamma\iDelta z.
\label{eA04}
\end{eqnarray}

\noindent
In (\ref{e212}) and (\ref{e213}) the affine connection $\iGamma$ is assumed 
to be independent of $z$, {\it i.e.}, $\iGamma=B$ and $C=0$.  
In this Appendix we have omitted suffices of $H^M_{\,N}$ and 
$\iGamma^M_{\,\,N5}$.

%%%%%%%%%%%%%%%%%%%%%%%%%%%%%% Appendix B %%%%%%%%%%%%%%%%%%%%%%%%%%%%%%%
\setcounter{equation}{0}
\addtocounter{section}{1}
\section*{Appendix \Alph{section}}

\indent

  Let us consider two paths on $M_4$
\begin{eqnarray}
x&\rightarrow&x+\iDelta_1 x\rightarrow (x+\iDelta_1 x)+\iDelta_2 x,
\label{eB01}\\
x&\rightarrow&x+(\iDelta_1 x+\iDelta_2 x).
\label{eB02}
\end{eqnarray}

\ \\
\begin{minipage}[t]{\minitwocolumn}
The difference $\Delta$ between two mapping functions is given by 
\end{minipage}
\hspace{\columnsep}
\begin{minipage}[t]{\minitwocolumn}
\begin{center}
\ \vspace{-1ex}\\
\setlength{\unitlength}{1mm}
\begin{picture}(80,30)(-10,-5)
 \put(30,5){\vector(0,1){11}}\put(30,16){\line(0,1){9}}
 \put(30,5){\vector(1,1){6}}\put(36,11){\line(1,1){4}}
 \put(40,15){\vector(-1,1){6}}\put(34,21){\line(-1,1){4}}
 \put(30,5){\circle*{1}}
 \put(40,15){\circle*{1}}
 \put(30,25){\circle*{1}}
 \put(19,27){$x+\iDelta_1 x+\iDelta_2 x$}
 \put(42,14){$x+\iDelta_1 x$}
 \put(29,1){$x$}
 \put(5,1){{\bf Fig.4.}}
\end{picture}
\vspace{-1ex}\\
\end{center}
\end{minipage}
\begin{eqnarray}
\Delta=H((x+\iDelta_1 x)+\iDelta_2 x,x+\iDelta_1 x)
       H(x+\iDelta_1 x,x)-H(x+(\iDelta_1 x+\iDelta_2 x),x).
\label{eB03}
\end{eqnarray}

\noindent
Substituting Eq.(\ref{e204}), {\it i.e.},
\begin{eqnarray}
H(x+\iDelta x,x)=1-\iGamma_\mu(x)\iDelta x^\mu
                 +\frac{1}{2}C_{\mu\nu}(x)\iDelta x^\mu\iDelta x^\nu+\cdots 
\label{eB04}
\end{eqnarray}

\noindent
into (\ref{eB03}), we have 
\begin{eqnarray}
\Delta=(\iGamma_\mu\iGamma_\nu-\iGamma_{\nu,\mu}
        -C_{\mu\nu})\iDelta_1 x^\mu\iDelta_2 x^\nu.
\label{eB05}
\end{eqnarray}

\noindent
If we choose $\iDelta_2 x^\mu=\alpha\iDelta_1 x^\mu\ (\alpha>0)$, two 
paths (\ref{eB01}) and (\ref{eB02}) become the same.  
In this case the difference $\Delta$ vanishes so that 
\begin{eqnarray}
C_{\mu\nu}=\frac{1}{2}(\iGamma_{\{\mu}\iGamma_{\nu\}}-\iGamma_{\{\nu,\mu\}}).
\label{eB06}
\end{eqnarray}

\noindent
Substituting this into (\ref{eB05}) we find 
\begin{eqnarray}
\Delta&=&\frac{1}{2}(\iGamma_{\mu,\nu}-\iGamma_{\nu,\mu}
                  +\iGamma_\mu\iGamma_\nu-\iGamma_\nu\iGamma_\mu)
                   \iDelta_1 x^\mu\iDelta_2 x^\nu\nonumber\\
      &=&\frac{1}{2}R_{\mu\nu}\iDelta_1 x^\mu\iDelta_2 x^\nu,
\label{eB07}
\end{eqnarray}

\noindent
where $R_{\mu\nu}$ is the Riemann curvature.  

  Let us now choose $\alpha=-1$, {\it i.e.}, 
$\iDelta_2 x^\mu=-\iDelta_1 x^\mu=-\iDelta x^\mu$.  
Then Eq.(\ref{eB03}) is reduced to 
\begin{eqnarray}
\Delta&=&H(x,x+\iDelta x)H(x+\iDelta x,x)-1 \nonumber\\
      &=&-\frac{1}{2}R_{\mu\nu}\iDelta x^\mu\iDelta x^\nu \nonumber\\
      &=&0,
\label{eB08}
\end{eqnarray}

\noindent
because $R_{\mu\nu}$ is antisymmetric with respect to $\mu$ and $\nu$.  
This shows that on $M_4$ there is no curvature corresponding to two paths: 
$x\rightarrow x+\iDelta x\rightarrow x$ and $x\rightarrow x$.

  In the discrete space, say $Z_N$, there is no case such that two paths 
$A\rightarrow B\rightarrow C$ and $A\rightarrow C$ become the same.  
So we calculate in $Z_2$ directly $\Delta$ defined by
\begin{eqnarray}
\Delta&=&1-H(+,-)H(-,+).
\label{eB09}
\end{eqnarray}

\noindent
Substituting (\ref{eA04}) into (\ref{eB09}) we find (\ref{e217}), 
the Riemann curvature of the third type.

%%%%%%%%%%%%%%%%%%%%%%%%%%%%% references %%%%%%%%%%%%%%%%%%%%%%%%%%%%%%%


\newpage
\noindent
\begin{thebibliography}{[00]}

\bibitem{Connes1} A. Connes, {\it The Interface of Math. \& Particle Phys.},
ed. D. Quillen, G. B. Segal and S.T. Tsou, Clarendon Press, Oxford(1990).\\
See also A. Connes, {\it Noncommutative Geometry}(Academic Press, Inc.,1994).
\bibitem{Connes2} A. Connes and J. Lott, Nucl. Phys.(Proc. Suppl.)
{\bf B18}(1990),~29.
\bibitem{NCG} D.Kaslter, {\it Introduction to Non-Commutative Geometry and 
Yang-Mills Model-Building, XIXth International Conference on Differential 
Geometric Methods in Theoretical Physics, Rapallo(Italy), June 1990, 
Springer Lect. Notes in Phys.} {\bf 375}(1990),~25.\\
R. Coquereaux, G. Esposito-Farese and G. Vaillant, Nucl. Phys. {\bf B353}
(1991),~689.\\
R. Coquereaux, R. H\"{a}\ss ling, N.A. Papadopoulos and F. Scheck, 
Int. J. Mod. Phys. {\bf A7}(1992),~2809.\\
A. Sitarz, Phys.Lett. {\bf B308}(1993),~311.\\
A.H. Chamseddine, G. Feldler and J. Frohlich, Phys. Lett. {\bf B296}
(1992),~109; Nucl. Phys. {\bf B395}(1993),~672.\\
K. Morita, Prog. Theor. Phys. {\bf 90}(1993),~219.\\
H.G. Ding, H.Y. Guo, J.M. Li and K. Wu, Z. Phys. {\bf C64}(1994),~512.\\
K. Morita and Y. Okumura, Prog. Theor. Phys. {\bf 91}(1994),~959; Phys. Rev.
{\bf D50}(1994),~1061.\\
Y. Okumura, Phys. Rev. {\bf D50}(1994),~1026; Prog. Theor. Phys. 
{\bf 92}(1994),~625.\\
S. Naka and E. Umezawa, Prog. Theor. Phys. {\bf 92}(1994),~189.
C.Y. Lee, D.S. Hwang and Y. Ne'eman, {\it ``BRST Quantization 
of Gauge Theory in Noncommutative Geometry: 
Matrix Derivative Approach'',hep-th/9512215}.\\
Y. Okumura, {\it ``BRST invariant Lagrangian of spontaneously broken 
gauge theories in noncommutative geometry'',hep-th/9603045}. 
Other related refs. are cited in this paper.
\bibitem{Konisi} G. Konisi and T. Saito, Prog. Theor. Phys. {\bf 95}
(1996),~657.\\
B. Chen, T. Saito, H.B. Teng, K. Uehara and K. Wu, Prog. Theor. Phys. {\bf 95}
(1996),~1173.
T. Saito and K. Uehara, {\it ``BRST Quantization of Gauge Theory on 
$M_4\times Z_2$ without Recourse to Noncommutative Geometry'',
hep-th/9607069}.\\
\bibitem{Chamseddine} A.H. Chamseddine, G. Felder and J. Frohlich, 
Commun. Math. Phys. {\bf 155} (1993),~205.
\bibitem{Kalau} W. Kalau, and M. Waltze, Preprint MZ-TH/93-38.
\bibitem{Kastler} D. Kastler, CPT-93/p2970.
\bibitem{Landi} G. Landi, N.A. Viet and K.C. Wali, Phys. Lett. {\bf B326}
(1994),~45.
\bibitem{Chen} B. Chen, T. Saito and K. Wu, Prog. Theor. Phys. {\bf B92}
(1994),~881.
\bibitem{Konisi2} G. Konisi, T. Saito and K. Wu, Prog. Theor. Phys. {\bf B93}
(1995),~621.
\bibitem{Brans} C.H. Brans and R.H. Dicke, Phys. Rev. {\bf 124} (1961),~925.

\end{thebibliography}
\end{document}